\documentclass[twoside]{article}
\usepackage{fleqn,espcrc2}

\usepackage{epsfig}

\newcommand{\be}{\begin{equation}}
\newcommand{\ee}{\end{equation}}
\newcommand{\bea}{\begin{eqnarray}}
\newcommand{\eea}{\end{eqnarray}}
\newcommand{\beqa}{\begin{eqnarray*}}
\newcommand{\eeqa}{\end{eqnarray*}}
\newcommand{\nn}{\nonumber}

\newcommand{\cD}{{\cal D}}

\title{Spintronics and Quantum Computing: Switching Mechanisms for
Qubits\thanks{To appear in Physica E, proceedings of the PASP2000 on the physics and
application of spin-related phenomena in semiconductors, Sendai, Japan, 2000.}}

\author{Michael N. Leuenberger\address[Basel]{Department of Physics and
Astronomy, University of Basel, \\
Klingelbergstrasse 82, CH-4056 Basel, Switzerland}, Daniel
Loss\addressmark[Basel]}

\begin{document}

\begin{abstract}
Quantum computing and quantum communication are remarkable examples of
new information processing technologies that arise from the coherent
manipulation of spins in
nanostructures.
We review our theoretical proposal for using electron spins in
quantum-confined nanostructures as qubits. We present single- and
two-qubit gate
mechanisms in laterally as well as vertically coupled quantum dots and
discuss the possibility to couple spins in quantum dots via exchange or
superexchange. In addition, we propose a new stationary wave switch,
which allows to perform quantum operations with quantum dots or
spin-1/2 molecules placed on a 1D or 2D lattice.
\end{abstract}

\maketitle

\section{Introduction}

Recent spin-related experiments with
electrons\cite{Prinz,Kikkawa,Fiederling,Ohno,Roukes,Ensslin}
have attracted much interest since the spin of the electron was shown to
reach very long dephasing times of the order of microseconds in
quantum-confined nanostructures\cite{Kikkawa,Fiederling,Ohno}, as well
as surprisingly long phase coherence lengths of up to $100\:\mu{\rm m}$
\cite{Kikkawa}. These achievements provide interesting possibilities for
finding novel mechanisms for information processing and information
transmission, such as
quantum computing\cite{Steane,Loss97,MMM2000} and spin-based devices for
conventional\cite{Prinz} computers. Not only that the fields of quantum
computing\cite{Steane,MMM2000} and quantum
communication\cite{MMM2000,Bennett00} could revolutionize
computing, but also the performance of quantum
electronic devices in conventional computers can be enhanced by the
electron spin,
e.g. spin-transistors (based on
spin-currents and spin injection), non-volatile memories,
single spin as the ultimate limit of information storage
etc.\cite{Prinz,Recher2}.
In Ref.\cite{Loss97} we have shown that
the electron spin is a most natural candidate for a qubit, which,
when located in quantum-confined structures such
as semiconductor quantum dots, atoms or molecules, satisfy all
requirements needed for a scalable quantum computer.
In particular, the Heisenberg exchange coupling between spins
in neighboring quantum dots
creates spin-entanglement  which is needed for qubit gates in quantum
computers, as well as  for producing mobile Einstein-Podolsky-Rosen (EPR)
pairs for quantum
communication\cite{BEL}.

In this paper we
review  our proposals for spin-based switching mechanisms for single-qubit
and two-qubit operations on  1D chains or 2D arrays of qubits; in
addition we present a new proposal for switching which might  be
interesting for
atomic systems.
A more extensive review of our recent work can be found in
Ref.~\cite{BEL} and references therein.

\section{Quantum Gate Operations with Coupled Quantum Dots}
\label{coupling}

One and two qubit gates are known to be sufficient to carry out any
quantum algorithm. For electron spins in nearby coupled quantum dots the
desired two qubit coupling is provided by a combination of Coulomb
interaction and the Pauli exclusion principle.

At zero magnetic field, the ground state of two
coupled electrons is a spin singlet,
whereas the first excited state in the presence of strong Coulomb
repulsion is usually a triplet. The remaining spectrum is separated
from these two states by a gap which is either given by the Coulomb
repulsion or the single particle confinement.
The low-energy physics of such a system can then be described by
the Heisenberg spin Hamiltonian
\begin{equation}\label{Heisenberg}
H_{\rm s}(t)=J(t)\,\,{\bf S}_1\cdot{\bf S}_2,
\end{equation}
where $J(t)$ is the exchange coupling  between
the two spins ${\bf S}_{1}$ and ${\bf S}_{2}$, and is given by the
energy
difference between the singlet and triplet states.
If we pulse the exchange coupling such that $\int dtJ(t)/\hbar =
J_0\tau_s/\hbar = \pi$ (mod $2\pi$), the associated unitary time
evolution $U(t) = T\exp(i\int_0^t H_{\rm s}(\tau)d\tau/\hbar)$
corresponds to the ``swap'' operator $U_{\rm sw}$ which
exchanges the quantum states of qubit 1 and 2 \cite{Loss97}. Having an
array of dots it is therefore possible to couple any two qubits.
Furthermore,
the quantum XOR gate, $U_{\rm XOR}$, can be constructed by applying
an appropriate sequence\cite{Loss97} of
``square-root of swap'' $U_{\rm sw}^{1/2}$ and single-qubit rotations
$\exp(i\pi S_i^z)$. Since $U_{\rm XOR}$ (combined with
single-qubit rotations) is proven to be a universal quantum
gate\cite{Barenco}, it can be used to assemble any quantum algorithm.
The study of universal quantum computation in coupled quantum dots (or
atoms etc.)
is thus essentially reduced to the study of single qubit rotations and
the {\it exchange mechanism}, in particular how the exchange coupling
$J(t)$ can be controlled experimentally.
Note that the switchable coupling mechanism described
below need not be restricted to quantum dots: the same
principle can be used in other systems, e.g. coupled atoms in a
Bravais lattice, supramolecular structures, or
overlapping shallow donors in semiconductors.

\begin{figure}[htb]
  \begin{center}
    \leavevmode
\epsfxsize=6cm
\epsffile{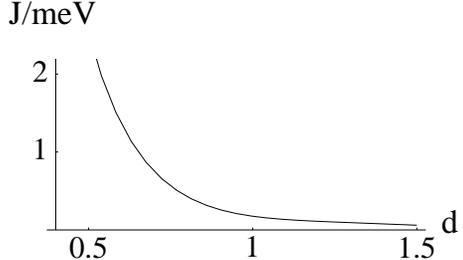}
  \end{center}
\caption{
Exchange coupling $J$ as a function of $d=a/a_B$ for $B=0$ and $c=2.42$,
where either the distance $a$ between the quantum dots (see
Sec.~\ref{lateral}) or the Bohr radius $a_B$ of the quantum dots (see
Sec.~\ref{2D}) is varied.}
\label{J(d)}
\end{figure}

\subsection{Laterally coupled quantum dots}
\label{lateral}

We first discuss a system of two laterally coupled quantum dots
defined by depleted regions in a 2DEG containing one (excess) electron
each\cite{Burkard}. The electrons are allowed to tunnel between the dots
(if the tunnel barrier is low) leading to spin correlations  via their
charge (orbital) degrees  of freedom.
We model the coupled system with the Hamiltonian
$H = H_{\rm orb} + H_{\rm Z}$, where $H_{\rm orb} = \sum_{i=1,2} h_i+C$
with
\bea
h_i & = & \frac{1}{2m}\left({\bf p}_i-\frac{e}{c}{\bf A}({\bf r}_i)
\right)^2+V({\bf r}_i), \nn\\
C & = & {{e^2}\over{\kappa\left| {\bf r}_1-{\bf r}_2\right|}}.
\label{hamiltonian}
\eea
Here, $h_i$ describes the single-electron dynamics in the 2DEG
confined to the $xy$-plane, with $m$ being the effective electron mass.
We allow for a  magnetic field ${\bf B}= (0,0,B)$ applied along the
$z$-axis that couples to the electron charge via the
vector  ${\bf A}({\bf r}) = \frac{B}{2}(-y,x,0)$, and to the
spin
via a Zeeman coupling term $H_{\rm Z}$.
The single dot confinement as well as the tunnel-coupling is modeled by
a quartic
potential,
$V(x,y)=\frac{m\omega_0^2}{2}\left(\frac{1}{4 a^2}\left(x^2-a^2
\right)^2+y^2\right)$,
which, in the limit $a\gg a_{\rm B}$,  separates  into two harmonic
wells  of
frequency $\omega_0$ where $2a$ is the interdot distance and
$a_{\rm B}=\sqrt{\hbar/m\omega_0}$
is the effective Bohr radius of a dot.
This choice for the potential is motivated by the experimental
observation\cite{tarucha} that
the low-energy spectrum of single dots is well described by a parabolic
confinement potential.
The (bare) Coulomb interaction between the two electrons is
described by $C$ where $\kappa$ denotes the dielectric constant of the
semiconductor. The screening length $\lambda$ in almost depleted regions
like few-electron quantum dots can be expected to be much larger than
the
bulk 2DEG screening length (about $40\,{\rm nm}$ for GaAs).
Therefore, $\lambda$ is large compared to the size of the coupled
system,
$\lambda\gg 2a\approx 40\,{\rm nm}$
for small dots, and we will consider the limit of unscreened
Coulomb interaction
($\lambda/a\gg 1$).
At low temperatures $kT_{B}\ll \hbar\omega_0$ we are allowed
to restrict our analysis to the two lowest orbital eigenstates of
$H_{\rm orb}$, leaving us with a symmetric (spin-singlet) and an
antisymmetric
(three triplets $T_{0}$, $T_{\pm}$) orbital state.
In this reduced (four-dimensional) Hilbert space, $H_{\rm orb}$
can be replaced by the effective Heisenberg spin Hamiltonian
Eq.~(\ref{Heisenberg}) where
the exchange coupling $J=\epsilon_{\rm t}-\epsilon_{\rm s}$ is
given by the difference between the triplet and
singlet energy. We make use of the analogy between atoms
and quantum dots (artificial atoms) and calculate
$\epsilon_{\rm t}$ and $\epsilon_{\rm s}$ with variational methods
similiar to the ones used in molecular physics.
With the Heitler-London approximation using single-dot groundstate
orbitals we find for 2D dots\cite{Burkard},
\bea
\label{J}
\lefteqn{J =
\frac{\hbar\omega_0}{\sinh\left(2d^2\,\frac{2b-1}{b}\right)}
\Bigg\{
\frac{3}{4b}\left(1+bd^2\right)+ c\sqrt{b}
}
\hspace{2cm}\\ \nn
 && \hspace{-2.8cm}  \times\left[e^{-bd^2} \, I_0\left(bd^2\right)
- e^{d^2 (b-1)/b}\, I_0\left(d^2\,\frac{b-1}{b}\right)\right]
\Bigg\},
\eea
where we introduce the dimensionless distance $d=a/a_{\rm B}$ and
the magnetic compression factor
$b=\sqrt{1+\omega_L^2/\omega_0^2}$,
where $\omega_L=eB/2m$ is the Larmor frequency.

${\rm I_0}$ denotes the zeroth Bessel function.
The first term in Eq.~(\ref{J}) comes from the confinement potential.
The terms
proportional to $c=\sqrt{\pi/2}(e^2/\kappa a_{\rm B})/\hbar\omega_0$ are
due to the Coulomb interaction $C$, where the exchange term appears with
a minus sign.
Note that typically $|J/\hbar\omega_0|\ll 1$ which makes the exclusive
use of ground-state single-dot orbitals in the Heitler-London ansatz
a self-consistent approach.
The most remarkable feature of $J(B)$ is the change of sign
from positive (ferromagnetic) to negative (antiferromagnetic), which
occurs at some finite $B$ over a wide
range of parameters $c$ and $a$.
This singlet-triplet crossing is caused by the long-range Coulomb
interaction and is therefore absent in the standard Hubbard model that
takes only into account short range interaction and, in the limit
$t/U\ll 1$,  is given by $J=4t^2/U>0$.
Large magnetic fields ($b\gg 1$) and/or large interdot distances ($d\gg
1$) reduce the overlap between the dot orbitals leading to an
exponential decay of $J$ contained in the $1/\sinh$ prefactor in
Eq.~(\ref{J}).
This exponential suppression is partly compensated
by the exponentially growing exchange term $\propto
\exp(2d^2(b-1/b))$. As a consequence, $J$ decays
exponentially as $\exp(-2d^2b)$ for large $b$ or $d$.
Thus, $J$ can be tuned through zero and then exponentially suppressed to
zero by a magnetic field in a
very efficient way (exponential switching is highly desirable to
minimize
gate errors). Further, working around the singlet-triplet crossing
provides a smooth exchange switching, requiring only small local
magnetic fields.
Qualitatively similar results are obtained\cite{Burkard} when we extend
the Heitler-London result by taking into account higher levels and
double
occupancy of the dots (using a Hund-Mullikan approach).
In the absence of tunneling ($J=0$) direct Coulomb interaction between
the electron charges can still be present. However the spins (qubit)
remain unaffected provided the spin-orbit coupling is sufficiently
small, which is the case for s-wave electrons in GaAs structures with
unbroken inversion symmetry.
Finally, we note that a spin coupling can also be achieved on a long
distance scale by using a cavity-QED scheme\cite{Imamoglu} or
superconducting leads to which the quantum dots are attached\cite{CBL}.

\subsection{Vertically coupled quantum dots}

We also investigated vertically coupled Quantum dots\cite{vertical}.
This kind of coupling can be implemented with multilayer self-assembled
quantum dots\cite{luyken} as well as with etched mesa
heterostructures\cite{austing}.

We model the vertical coupled dot system by a potential $V=V_l+V_v$
where $V_{l}$ describes the parabolic lateral confinement and $V_{v}$
models the vertical dot coupling assumed to be a quartic potential
similar to the one introduced above for the lateral coupling. We allow
for different dot sizes $a_{{\rm
B}\pm}=\sqrt{\hbar/m\alpha_{0\pm}\omega_z}$ with $\omega_{z}$ being the
vertical confinement, implying an effective Bohr
radius $a_{\rm B}=\sqrt{\hbar/m\omega_z}$ and
 a dimensionless interdot distance $2d = 2a/a_{\rm B}$. By applying an
in-plane electric field $E_\parallel$ an
interesting new switching mechanism arises. The dots are shifted
parallel to the field
  by $\Delta x_\pm =E_\parallel/E_0\alpha_{0\pm}^2$, where
$E_0=\hbar\omega_z/ea_B$.
Thus, the larger dot is shifted a greater distance $\Delta x_{-}>\Delta
x_{+}$
 and so the mean distance between the electrons grows as
 $d'=\sqrt{d^2+A^2(E_\parallel/E_0)^2}>d$, taking
 $A=(\alpha_{0+}^2-\alpha_{0-}^2)/2\alpha_{0+}^2\alpha_{0-}^2$.
Since the exchange coupling $J$ is exponentially sensitive to the
interdot distance $d'$ (see Eq. (\ref{J}))
we have another exponential switching
mechanism for quantum gate operations at hand.

\begin{figure}[hbt]
  \begin{center}
    \leavevmode
\epsfxsize=8cm
\epsffile{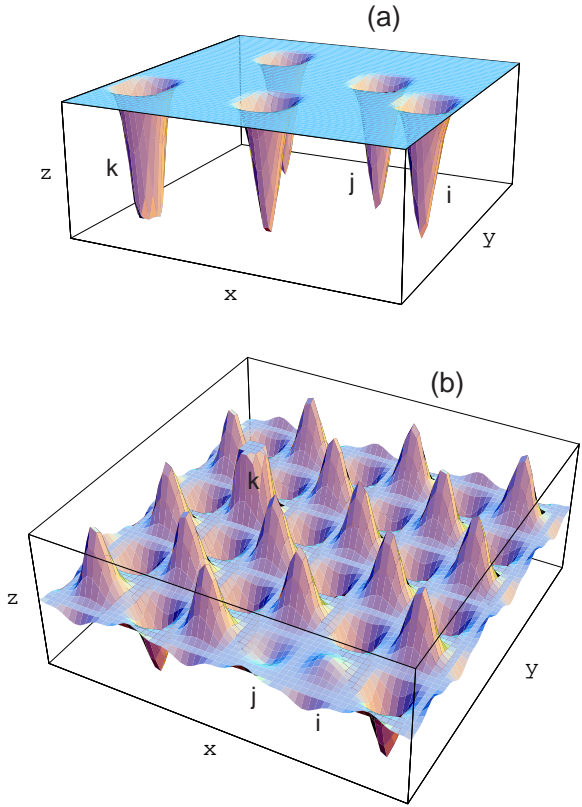}
  \end{center}
\caption{
Potentials for the (a) 1D chain and (b) 2D array of quantum dots that
are manipulated by a (a) 1D or (b) 2D standing waves produced by
gates (see Fig.~\ref{sandwich}) or lasers. All exchange interactions
$J_{pq}=0$ between qubit $p$ and $q$, except $J_{ij}\ne 0$ for a given
switching time $\tau_s$. The potential for qubit $k$ is deformed such
that the electron spin contained in it is closer to some magnetized/high-$g$
layer, allowing a  single-spin manipulation (see text).}
\label{standing_wave}
\end{figure}

\subsection{Coupling two spins by superexchange}

There is a principal problem if one wants to couple two ``extended"
dots  whose energy
levels are closely spaced (i.e. smaller than $k_BT$), as would
be the case
if there is a sizable distance between the
two
confined qubits before the barrier is lowered.
 In this case, the
singlet-triplet splitting
becomes
vanishingly small, and it would not take much excitation energy to get
states which are not
entangled at all. In other words, the adiabatic switching
time\cite{Burkard} which is proportional to the
inverse
level spacing becomes arbitrarily large. A better scenario for coupling
the two
spin-qubits is to
make use of a superexchange mechanism to obtain a Heisenberg
interaction\cite{Loss97,Recher}. Consider three aligned quantum
dots where the middle dot is empty and
so small that only its lowest
levels  will be occupied by
1 or 2 electrons in a virtual
hopping process.
The left
and right dots can be much larger
but still small enough such that the
Coulomb charging energies $U_{L}\approx U_{R}$  are high enough
(compared to $k_BT$) to suppress any double occupancy. Let
us assume now that the middle
dot has  energy levels higher than the
ground states of right
and left
dots, assumed to be approximately the same. These levels
include single particle energy (set to zero) and
Coulomb charging energy $N^2e^2/2C$, with $N$ the number of electrons
and C the capacitance of the middle dot,
and thus  the ground state energy of the middle dot is $0$ when
empty, $\epsilon=e^2/2C$  for one electron, and $4\epsilon$ for 2
electrons.
The tunnel coupling between the dots is denoted by $t_{0}$. Now,
there
are two types of virtual processes possible which couple the spins but
only one is dominant.
First,
the electron of the left (right) dot hops on the middle dot, and then
the electron from the
right
(left) dot hops on the {\it same} level on the middle dot, and thus, due
to the Pauli principle, the
two electrons on the middle dot form a singlet, giving the desired
entanglement. And then they
hop
off again into  the left and right dots, respectively. (Note that U must
be
larger than $k_{B}T$,
otherwise
real processes involving 2 electrons in the left or right dot will be
allowed). It is not
difficult to
see that this virtual process leads to an effective Heisenberg exchange
interaction with exchange constant $J=4t_{0}^4/4\epsilon^3$, where the
virtual energy denominators follow the sequence $1/\epsilon\rightarrow
1/4\epsilon\rightarrow
1/\epsilon$.

In the second type of virtual process the left (right)
electron hops via the middle dot into
the right (left) dot and forms there a singlet, giving
$J=4t_{0}^4/U_{R}\epsilon^2$.
However, this process has
vanishing weight because there are also many nearby states available in
the outer dots
for which
there is no spin correlation required by the Pauli principle. Thus, most
of the virtual
processes,
for which we have 2 electrons in the left (right) dot, do not produce
spin
correlations, and
thus we
can neglect these virtual processes of the second type altogether.

\section{Single-Spin Rotations}

In order to perform one qubit gates single-spin rotations are required.
This is done by exposing  a single spin to a time-varying Zeeman
coupling
 $(g\mu_B {\bf S}\cdot {\bf B})(t)$ \cite{Burkard},
 which can be controlled through
 both the magnetic field ${\bf B}$ and/or the g-factor $g$.
We have proposed a number of possible
implementations\cite{Loss97,Burkard,MMM2000,BEL} for spin-rotations:
Since only relative phases between qubits are relevant we can apply  a
homogeneous ${\bf B}$-field rotating all spins at once. A local change
of the Zeeman coupling is then possible by changing the Larmor frequency
$\omega_{L}=g\mu_{B}B/\hbar$. The equilibrium position of an electron
can be changed through electrical gating, therefore if the electron
wavefunction is pushed into a region with different magnetic field
strength or different (effective) g-factor, the relative phase of such
an electron then becomes $\phi = (g'B'-gB)\mu_B\tau/2\hbar$. Regions
with an increased magnetic field can be provided
 by a magnetic (dot) material  while an effective magnetic field can be
produced e.g.\ with dynamically polarized nuclear spins (Overhauser
effect)\cite{Burkard}.

Alternatively one can use electron-spin-resonance (ESR) techniques
\cite{Burkard}  to perform single-spin rotations, e.g. if we want to
flip a certain qubit (say from $|\uparrow\rangle$ to
$|\downarrow\rangle$) we apply an ac-magnetic field perpendicular to the
$\uparrow$- axis that matches the Larmor frequency of that particular
electron. Due to paramagnetic resonance the spin can flip.

\begin{figure}[htb]
  \begin{center}
    \leavevmode
\epsfxsize=8cm
\epsffile{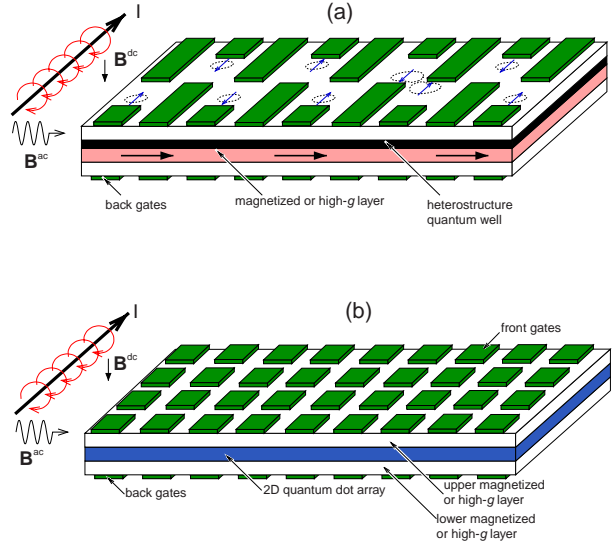}
  \end{center}
\caption{
(a) 1D chain (b) 2D array of quantum dots sandwiched by two (a) 1D
chains (b) 2D arrays of gates, which can be contacted (a) laterally (b)
along the axis perpendicular to the $xy$-plane (not shown). The
$B$-fields and magnetized/high-$g$ layers can be used for single-qubit
switching (see text).}
\label{sandwich}
\end{figure}

\section{1D and 2D Stationary Wave Switch}
\label{2D}

In this section we propose a new method to perform one- and two-qubit
operations on a 1D linear chain and a 2D array of quantum dots, which
promises to be extendable to spin-1/2 molecules or even atoms. We
define the 2DEG to lie in the $xy$-plane. The initial configuration (IC)
of the $\cD$ lattice of quantum dots ($\cD=1D,2D$), in which all the
isotropic exchange couplings $J_{pq}$ between sites $p$ and $b$ are
zero, is represented by a $\cD$ stationary wave ($\cD$SW). The 1DSW
refers to the sinusoidal arrangement of the 1D chain of quantum dots
controlled by lateral gates on either side of the chain in the
$xy$-plane as shown in Fig.~\ref{standing_wave} (a), whereas the 2DSW
denotes the 2D potential (see  Fig.~\ref{standing_wave} (b)) produced by
the front and back gates that are positioned above and below the 2D
quantum dot array (see Fig.~\ref{sandwich}), in which the dots are
trapped. In this way no split gates are needed. Instead of using
gates, it should be possible to produce the $\cD$SW by means of, say,
x-ray lasers\cite{x-ray}, which would confine electrostatically the
quantum dots in the
IC. X-ray lasers could also be used to distort a $\cD$ crystal lattice
made of spin-1/2 molecules or atoms in order to obtain the IC. As these
molecules are neutral in charge, the stationary electrostatic fields
produce locally electric dipoles within these molecules, leading to
a periodic displacement of the electron spin (qubit) of interest.

In order to perform qubit operations, the $\cD$SW must be varied. For
operations on a single quantum dot in  1D or 2D the corresponding
extremum at which the quantum dot is located must be enhanced such that
the qubit is displaced into the magnetized or high-$g$ layer above or
below the 2D quantum dot array (see Fig.~\ref{sandwich} and the previous
section). The coupling between two quantum dots is achieved by reducing
the corresponding two pair of gates, which in 1D decreases the distance
$a$ between the two quantum dots, and in 2D locally suppresses the
extrema of the two quantum dots. $J(d)$ is given by Eq.~(\ref{J})
because two nearby quantum dots of the $\cD$SW shown in
Fig.~\ref{standing_wave} can be modelled by a quartic potential of the
form $V(x,y)=\frac{m\omega_0^2}{2}\left(\frac{1}{4
a^2}\left(x^2-a^2\right)^2+y^2\right)$. While in 1D the distance $a$ is
changed (see Sec.~\ref{lateral}), in 2D the potential height of $V(x,y)$
is altered, which corresponds to a change of the harmonic oscillator
frequency $\omega_0$. This in turn results in a variation of the Bohr
radius $a_B=\sqrt{\hbar/m\omega_0}$. For small magnetic fields $B\ll
2m\omega_0/e$, i.e. $b\approx 1$, the magnetic compression factor $b$ is
nearly independent of $\omega_0$. Then $J(d)$ in Eq.~(\ref{J}) becomes
a function of $d(a)$ in 1D or a function of $d(a_B)$ in 2D (see
Fig.~\ref{J(d)}). Thus, by varying $a$ or $\omega_0$ the exchange
$J$ can be switched on or off, see Fig.~\ref{J(d)}.


\begin{thebibliography}{00}

\bibitem{Prinz}
G. Prinz, Phys. Today {\bf 45}(4), 58 (1995);
G. A. Prinz, Science {\bf 282}, 1660 (1998).

\bibitem{Kikkawa}
D.D. Awschalom and J.M. Kikkawa, Phys.\ Today {\bf 52}(6), 33 (1999).

\bibitem{Fiederling} 
R. Fiederling  {\it et al.},
Nature {\bf 402}, 787 (1999).

\bibitem{Ohno}
Y. Ohno  {\it et al.},
Nature {\bf 402}, 790 (1999).

\bibitem{Roukes} F.G. Monzon and M.L. Roukes, J. Magn. Magn. Mater.
{\bf 198}, 632 (1999).

\bibitem{Ensslin}
S. L\"uscher {\it et al.},
cond-mat/0002226.

\bibitem{Steane}
A. Steane,
Rep. Prog. Phys. {\bf 61}, 117 (1998).


\bibitem{Loss97}
D. Loss and D.P. DiVincenzo,
Phys.\ Rev.\ A {\bf 57}, 120 (1998); cond-mat/9701055.


\bibitem{MMM2000}
D.P. DiVincenzo and D. Loss,
J. Magn. Magn. Mater. {\bf 200}, 202 (1999).

\bibitem{Bennett00} C. H. Bennett and D. P. DiVincenzo,
Nature {\bf 404}, 247 (2000).

\bibitem{Recher2} P.~Recher, E.~V.~Sukhorukov, D.~Loss,
Phys.~Rev.~Lett. {\bf 85}, 1962 (2000).


\bibitem{BEL}
G. Burkard, H.-A. Engel, and D. Loss,
to appear in Fortschritte der Physik, special issue on
\textit{Experimental Proposals for Quantum Computation}, eds. S.
Braunstein and K.L. Ho; cond-mat/0004182.

\bibitem{tarucha}
S. Tarucha {\it et al.},
Phys.\ Rev.\ Lett.\ {\bf 77},  3613  (1996).

\bibitem{Barenco}
A. Barenco {\it et al.},  Phys.\ Rev.\ A {\bf 52}, 3457 (1995).

\bibitem{Burkard}
G. Burkard, D. Loss, and D. P. DiVincenzo,
Phys.\ Rev.\ B {\bf 59}, 2070 (1999).

\bibitem{Imamoglu} A. Imamoglu,
{\it et al.},
Phys.\ Rev.\ Lett.\ {\bf 83}, 4204 (1999).

\bibitem{CBL}
M.-S. Choi, C. Bruder, and D. Loss, to appear in Phys.
Rev. B; cond-mat/0001011.

\bibitem{vertical}
G. Burkard, G. Seelig, and D. Loss; Phys.\ Rev.\ B {\bf 62}, 2581
(2000).



\bibitem{luyken}
R.~J. Luyken
{\em et al.},
preprint.

\bibitem{austing}
D.~G. Austing
{\em et al.},
Physica \ B {\bf 249-251}, 206 (1998).


\bibitem{Recher}
P. Recher, D. Loss, and J. Levy,
cond-mat/0009270.

\bibitem{x-ray} K.-D. Liss {\em et al.},
Nature (London) {\bf 404}, 371 (2000);
M.~Tegze {\em et al.}, Nature (London) {\bf 407}, 38 (2000).
These papers demonstrate pulsed x-ray lasers with 100 picoseconds pulse length
and wave lengths down to 10 pm.
For the present purpose it would be desirable  to have
x-ray lasers available that can be operated in a cw mode.
\end{thebibliography}
\end{document}